\documentclass{article}
\usepackage{spconf,amsmath,graphicx}
\usepackage{booktabs}
\usepackage{amssymb}
\usepackage[font=small,skip=4pt]{caption} 
\usepackage{xcolor}
\usepackage{hyperref}
\usepackage{enumitem}

\usepackage{eso-pic, rotating, graphicx}

\title{Improved Normalizing Flow-Based Speech Enhancement using an\\ All-pole Gammatone Filterbank for Conditional Input Representation}
\name{Martin Strauss, Matteo Torcoli, Bernd Edler}
\address{International Audio Laboratories Erlangen, a joint institution of the Friedrich-Alexander-Universität \\Erlangen-Nürnberg (FAU) and Fraunhofer IIS, Germany}
\copyrightnotice{978-1-6654-7189-3/22/\$31.00~\copyright2023 IEEE}
\begin{document}
%\ninept
%
\maketitle

\AddToShipoutPicture*{\put(600,750){\rotatebox{270}{\scalebox{1}{\footnotesize \copyright  Copyright 2023 IEEE. Published in the 2022 IEEE Spoken Language Technology Workshop (SLT) (SLT 2022), scheduled for 19-22 January 2023 in Doha, Qatar. Personal use of this material is permitted. However, permission to }}}} 
\AddToShipoutPicture*{\put(590,750){\rotatebox{270}{\scalebox{1}{\footnotesize  reprint/republish this material for advertising or promotional purposes or for creating new collective works for resale or redistribution to servers or lists, or to reuse any copyrighted component of this work in other works, must be }}}}
\AddToShipoutPicture*{\put(580,750){\rotatebox{270}{\scalebox{1}{\footnotesize  obtained from the IEEE. Contact: Manager, Copyrights and Permissions / IEEE Service Center / 445 Hoes Lane / P.O. Box 1331 / Piscataway, NJ 08855-1331, USA. Telephone: + Intl. 908-562-3966.}}}}

\begin{abstract}
Deep generative models for Speech Enhancement (SE) received increasing attention in recent years. %The goal is to recover clean speech from signals degraded by background noise. 
The most prominent example are Generative Adversarial Networks (GANs), while normalizing flows (NF) received less attention despite their potential. %The aim of this  paper is to shed further light on flow-based SE.
Building on previous work, architectural modifications are proposed% to increase its performance and the efficiency of the training process
, along with an investigation of different conditional input representations. % Maybe a reason why?
Despite being a common choice in related works, Mel-spectrograms demonstrate to be inadequate for the given scenario. Alternatively, a novel All-Pole Gammatone filterbank (APG) with high temporal resolution is proposed. Although computational evaluation metric results would suggest that state-of-the-art GAN-based methods perform best, a perceptual evaluation via a listening test indicates that the presented NF approach (based on time domain and APG) performs best, especially at lower SNRs. On average, APG outputs are rated as having \textit{good} quality, which is unmatched by the other methods, including GAN.
\end{abstract}
\section{Introduction}
\label{sec:intro}

Speech Enhancement (SE) aims to improve the quality of speech degraded by disturbing background noise \cite{Loizou2007}. Therefore, it plays a vital role as a front end for automatic speech recognition systems \cite{Pandey2021, ASRROB, ESPNET} and far-field speech processing \cite{Haeb-Umbach}.
%It has a variety of applications, including automatic speech recognition \cite{Pandey2021}, speech coding \cite{zhao2019convolutional}, hearing aids \cite{Park2020}, and broadcasting~\cite{torcoli2021ibc}.
SE was investigated extensively and approaches based on Deep Neural Network (DNN) largely overtook traditional techniques like spectral substraction \cite{Gustafsson2001}, Wiener filter \cite{Jingdong2006} or subspace methods \cite{Klein2002}.
Most commonly, a separation mask is estimated by minimizing a distance metric to extract the clean speech components in Time-Frequency (TF) domain \cite{Koizumi2021, Wang2017} or a learned subspace \cite{Luo2019ConvTasNetSI}.
Still, in recent years there has been an increasing interest in generative approaches trying to outline the probability distribution of speech signals. The most prominent examples include Generative Adversarial Networks (GANs) \cite{fu21_interspeech, hifigan2}, Variational Autoencoders (VAE) \cite{Leglaive2019}, autoregressive models \cite{Qian2017} and diffusion probabilistic models \cite{lu2022conditional}. GAN-based architectures stand out in their performance. For instance, MetricGAN+~\cite{fu21_interspeech} and HiFi-GAN-2~\cite{hifigan2} are the respective successors of adversarialy trained DNNs for SE. MetricGAN+ is directly optimized on PESQ \cite{pesq} or STOI \cite{stoi}, reporting high values in the corresponding metrics at the output. HiFi-GAN-2 is pretrained in a discriminative way, followed by adversarial optimization to improve perceptual quality. Although the presented results are genuinely impressive, GANs in general are known to be difficult to train in a stable manner and tend to suffer from mode collapse \cite{MetzPPS17}. \\
Diffusion probabilistic models are a recent example of generative models where the transformation from Gaussian noise to clean input is learned by a diffusion process. Lu et al.~\cite{lu2022conditional} were the first to apply this approach to SE, restoring clean speech by conditioning the process on noisy speech. They show a leading performance in time-domain generative models and promising generalization in mismatched conditions. Still, sampling from a diffusion process is rather slow and computationally expensive~\cite{song2021denoising}.
Normalizing Flows (NFs) \cite{Papamakarios2019NormalizingFF} are another generative modelling technique. They are trained by maximizing the likelihood of the data directly, making them easy and stable to train. Despite increasing success in fields like computer vision \cite{DinhSB17} or speech synthesis \cite{Prenger2019}, their application in SE has received less attention. Nugraha et al.~\cite{Nugraha2020} applied NFs  in combination with a VAE to learn a deep speech prior to be combined with a SE algorithm of choice. In contrast, Strauss et al.~\cite{strauss2021} used NFs to learn the mapping from Gaussian noise to clean speech conditioned on a noisy speech sample entirely in time domain. While outperforming the results of other time-domain GAN-based methods, the overall performance evaluated on computational metrics lag behind comparable TF domain approaches.\\
%  The choice of conditional signal was is not fully investigated. Commonly,  mel spect--> beispiele -->
Building on previous work, the aim of this paper is to give further insights on NF-based SE. We improve the architecture of \cite{strauss2021} by a simple double coupling scheme to ensure that the entire input signal is processed in one flow block. Further, different input representations for the conditional noisy input signal are considered. Our experiments show that despite the fact that Mel-spectrograms are a common choice for conditional signal representation in related fields, like neural vocoders~\cite{Prenger2019,stylemelgan}, they are inadequate for our scenario. Alternatively, the usage of a Bark-spaced All-Pole Gammatone filterbank (APG) \cite{Lyon_all-polemodels} is proposed. Similar to Mel this design makes use of a perceptually motivated filterbank to mimic the human auditory system and reduce the dimensionality of the filter output compared to a standard Short-Time Fourier Transform (STFT). At the same time temporal resolution with the design of this filterbank is increased, which  overcomes the limitations of a standard Mel-spectrogram. Perceptual evaluation via a listening test indicates that the presented NF approach (based on time domain and APG) performs better than state-of-the-art GAN-based methods, especially at lower SNRs, even though this is not reflected by computational evaluation metrics.

\section{Normalizing flow-based speech enhancement}

Let's define two $D$ dimensional random variables $\textbf{x} \in \mathbb{R}^{D}$ and $\textbf{z} \in \mathbb{R}^{D}$. A NF is defined by a differentiable function $f$ with differentiable inverse allowing a bijective transformation between the two random variables \cite{Papamakarios2019NormalizingFF} , i.e.,
\begin{equation}
	\textbf{x} = f(\textbf{z}),  \hspace{1cm} \textbf{z} = f^{-1}(\textbf{x}).
\end{equation}
The invertability of $f$ ensures that the random variable $\textbf{x}$ is defined by a given probability distribution and can be computed by a change of variables, i.e.,
\begin{equation}
	p_x(\textbf{x}) = p_z(\textbf{z}) \left\lvert \det \left( J(\textbf{x}) \right) \right\rvert ,
\end{equation}
where ${J(\textbf{x})=\partial \textbf{z}/ \partial \textbf{x}}$ is the Jacobian containing all first order derivatives. Since $f$ is invertible, this holds true also for a sequence of functions $f_{1:T}$, i.e.,
\begin{equation}
	\textbf{x} = f_1 \circ f_2 \circ \cdots \circ f_T(\textbf{z}).
\end{equation}
Let us now introduce a single channel noisy speech signal $\textbf{y} \in \mathbb{R}^{N}$ with sequence length $N$, obtained by the summation of a clean speech utterance $\textbf{x}\in \mathbb{R}^{N}$ and background noise $\textbf{n}\in \mathbb{R}^{N}$:
\begin{equation}
	\textbf{y}=\textbf{x}+\textbf{n}.
\end{equation}
Moreover, $\textbf{z} \in \mathbb{R}^{N}$ is defined to be sampled from a Gaussian distribution with zero mean and unit variance, i.e.,
\begin{equation}
	\textbf{z} \sim \mathcal{N}(\textbf{z}| 0,\textbf{I}).
\end{equation}
The aim of NF-based SE is now to outline the conditional probability distribution $p_x(\textbf{x}|\textbf{y})$ by a DNN with parameters $\theta$. Hence, the overall training objective is described by a maximization of the log-likelihood, i.e., 
\begin{equation}
	\log p_x(\textbf{x}|\textbf{y}; \theta) = \log p_z(f^{-1}_{\theta}(\textbf{x})|\textbf{y}) + \log \left\lvert \det \left( J(\textbf{x}) \right) \right\rvert.
\end{equation}
Inverting the learned network, a noise example sampled from $p_z(z)$ is conditioned on a noisy speech utterance and mapped back to the distribution of clean speech utterances resulting in an enhanced speech output.

\section{Proposed methods}
\subsection{Model architecture}
\begin{figure}[htb]
	\resizebox{0.47\textwidth}{!}{%
		\centering
		\begin{minipage}[b]{\textwidth}% 0.88
			\includegraphics[width=1\textwidth]{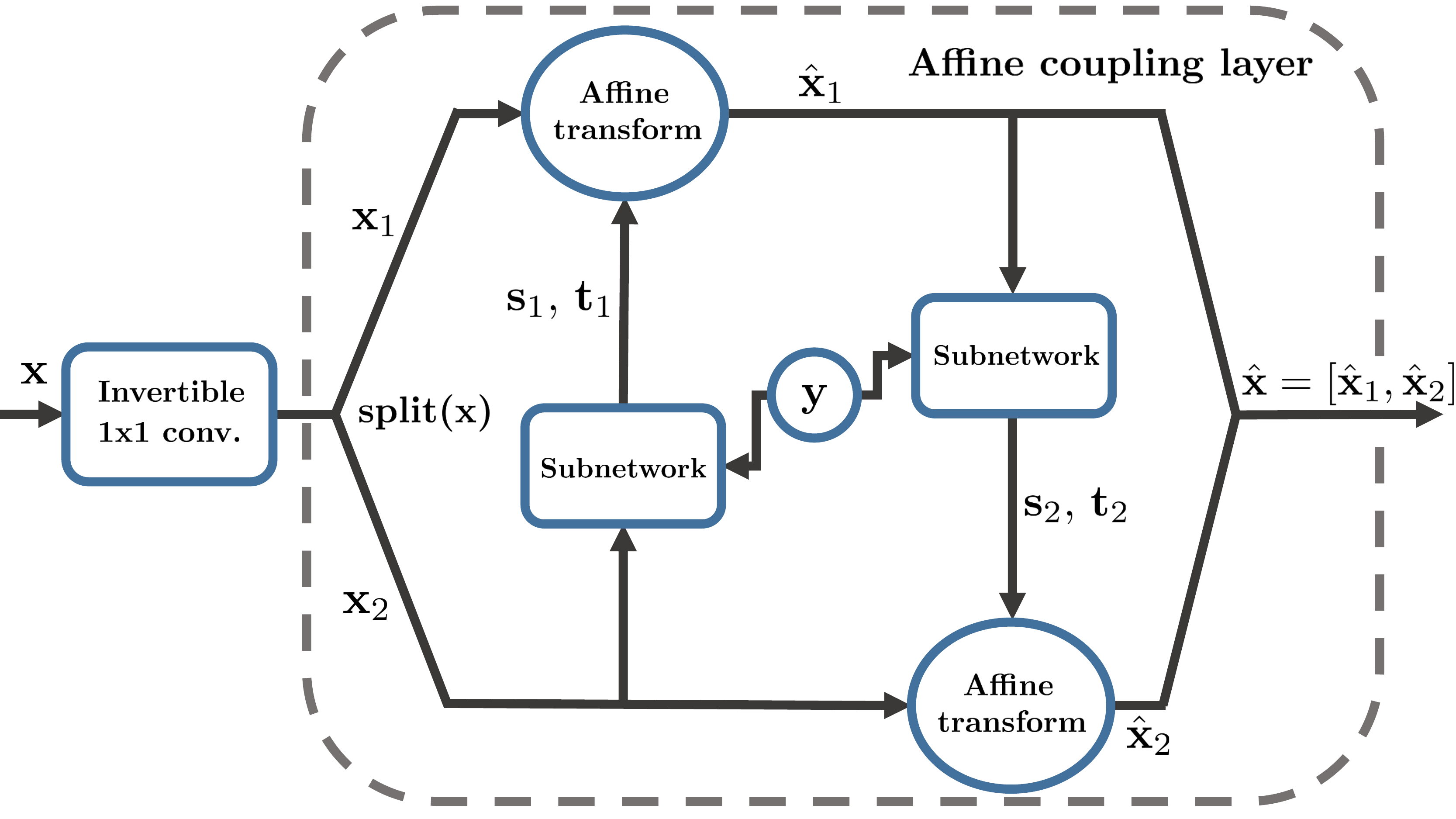}% mushra
	\end{minipage}}
	\caption{Double coupling scheme. Before entering the affine coupling layer, the subsampled input $x$ passes through the invertible 1x1 convolution. The conditional input $y$ serves as input to both subnetworks. For single coupling $x_2$ is passed unchanged through the identity function essentially leading to $\hat{x}_2 = x_2$.}
	%\vspace{-0.5cm}
	\label{fig:dc}
\end{figure}

The model used in the experiments builds upon \cite{strauss2021}. This network consists of a sequence of so-called flow blocks to transform the input clean speech utterance to Gaussian noise. One flow block (Figure~\ref{fig:dc}) contains a combination of an invertible 1x1 convolutional layer \cite{glow2018} and an affine coupling layer \cite{DinhSB17}. Similar to the Glow \cite{glow2018} network, each block processes multiple channels of the input signal at once. Therefore, the input $x \in \mathbb{R}^{1 \times N}$ with sequence length $N$ is subsampled by a factor $G$ to create a multichannel signal $x \in \mathbb{R}^{G \times (N/G)}$. After the invertible convolutional layer, the input is separated into two halves along the channel dimension with one part being provided to the subnetwork inside the coupling layer to learn affine transformation parameters $s$ and $t$ for the second half. The transformed signal is concatenated with the unchanged second part and serves as an input for the next block. This operation is invertible, ensuring that the network is invertible overall, although the subnetwork inside the coupling layer estimating the affine parameters does not need to be invertible. 

%With every coupling block only one half of the input signal is transformed, while the second half remains unchanged. 
To increase the capacity of each block, a double coupling scheme inspired by \cite{ardizzone2018analyzing} was implemented where the output of the affine transformation is reused as an input to calculate the affine parameters for the second part, i.e.,

\begin{equation}
	\begin{split}
		\hat{x}_1 = s_1(x_2) \odot x_1 + t_1(x_2), \\
		\hat{x}_2 = s_2(\hat{x}_1) \odot x_2 + t_2(\hat{x}_1),
	\end{split}
\end{equation}
where the input $x$ is separated into $x_1$ and $x_2$ and $s_1$, $t_1$, as well as $s_2$, and $t_2$ are estimated by respective subnetworks. The output is concatenated, i.e., $\hat{x}=\left[ \hat{x}_1,\hat{x}_2 \right] $, and passed to the next flow block. This procedure is illistrated in Figure~\ref{fig:dc}. 

Similar to Waveglow \cite{Prenger2019} the subnetwork used in this paper is a stack of dilated convolutions with skip connections applied to the input signal. The conditional signal is also subsampled by the factor $G$, before being processed by a single convolutional layer and introduced to each layer of the subnetwork by a gated activation, i.e. a combination of a tanh and sigmoid function, as proposed for WaveNet~\cite{wavenet}.

\subsection{Conditional input representations}
\begin{figure}[htb]
	\centering
	\resizebox{0.45\textwidth}{!}{%
		\centering
		\begin{minipage}[b]{\textwidth}% 0.88
			\includegraphics[width=1\textwidth]{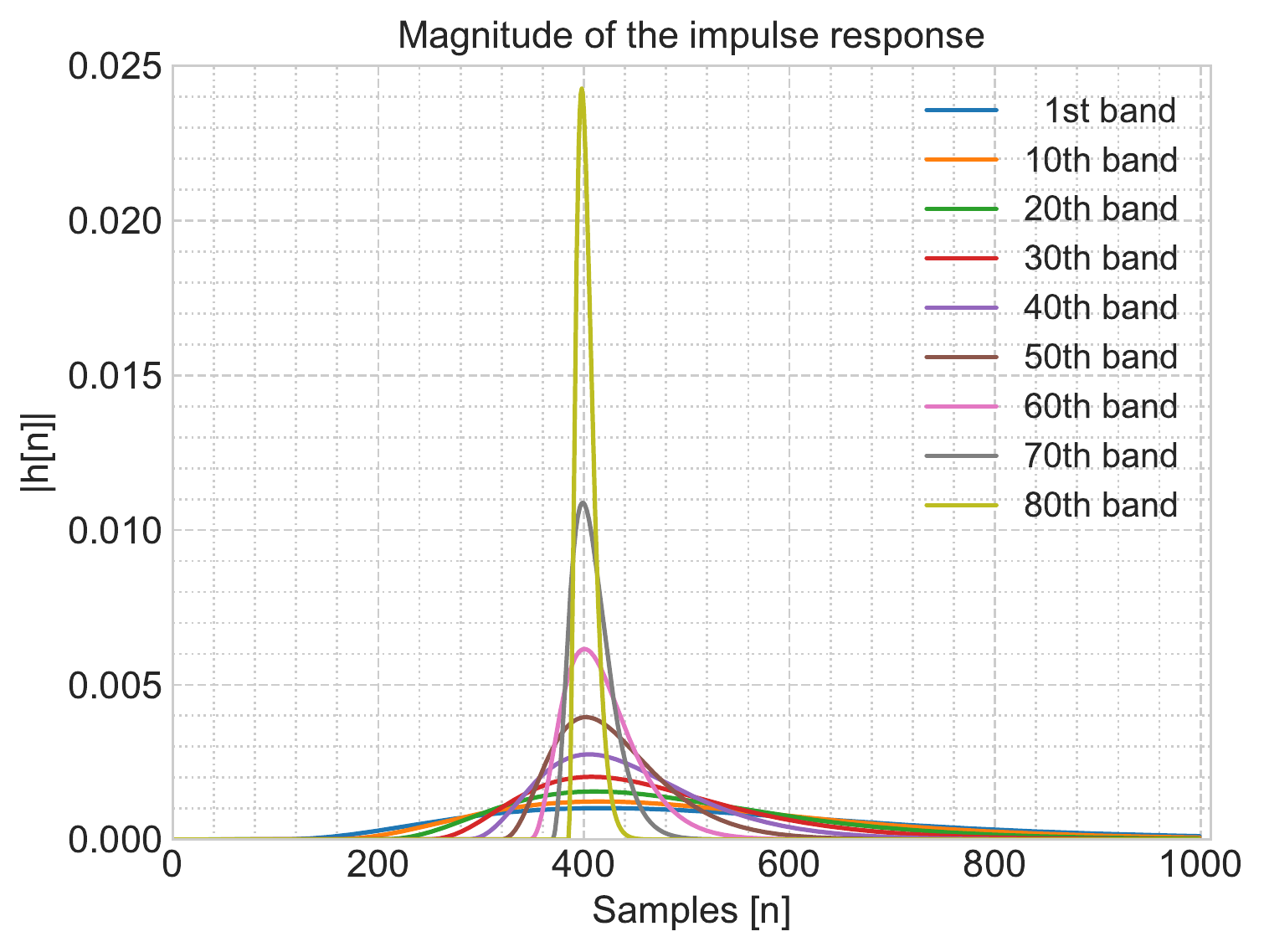}% mushra
	\end{minipage}}
	\caption{Magnitudes of the filter responses of the All-Pole Gammatone filterbank (APG). The filter outputs are delay compensated. For better visibility, only every 10th band is displayed.
		(Best viewed in colors)}
	\vspace{-0.5cm}
	\label{fig:apg}
\end{figure}

Next to the original time domain as representation of the conditional signal, experiments with additional variations are conducted. The Mel-spectrogram is a common choice for conditional representation and works well, e.g., in neural vocoders~\cite{Prenger2019, stylemelgan}. Each Mel spectral coefficient is, as usual, computed from FFT magnitudes by multiplication with a triangular spectral weighting function and summation. For instance, in ~\cite{Prenger2019} an FFT length of 1024 samples (at $f_s=22$\,kHz) was chosen to obtain a frequency resolution appropriate for achieving sufficient accuracy of the lowest Mel coefficients. This, however, leads to a time resolution which is much below that of the human auditory system at higher frequencies, so that fine temporal structures are not sufficiently represented. Further, the time frames are up-sampled to match the time input dimension using a transposed convolution layer. This step is rather redudant, since the  low number of time frames needs to be brought up to full time resolution again without adding further information. Consequently, in initial experiments using a Mel representation phase/time shifts occured in the enhanced output, which were accounted to the low temporal resolution in the conditional input. Moreover, at enhancement the only useful information provided to the network is the noisy conditional signal and corrupted time frames potentially do not provide sufficient information.\\
\indent Therefore, we investigated the use of a specifically designed complex-valued All-Pole Gammatone filterbank (APG). Motivated by human hearing, the center frequencies have constant distances on the Bark scale with increasing bandwidth at increasing frequencies, proportional to the Bark bandwidths. The IIR filters operate directly on the time domain input signal with cascades of first order complex valued stages. Thus, filter outputs at the input sampling rate are obtained. Although we use only the output magnitudes as conditional input, their temporal resolutions are only limited by the extent of the impulse responses, which are longer for the narrow bands with lower center frequencies, but relatively short for the wider bands with higher center frequencies as can be seen in Figure~\ref{fig:apg}.

\section{Experimental setup}
\subsection{Dataset}
For the experiments we consider the commonly used Voice-Bank-DEMAND dataset \cite{valentini2016}. It includes 30 speakers separated into 28 for training and 2 for testing. The dataset items consist of speech samples from the VoiceBank corpus~\cite{voicebank} corrupted with noise items from the DEMAND database~\cite{demand} and artificially generated speech shaped and babble noise. The items are mixed together with Signal-to-Noise-Ratios (SNRs) of 0, 5, 10, and 15\,dB for training. For testing, SNRs of 2.5, 7.5, 12.5, 17.5\,dB are used. In our experiments, one male and one female speaker are taken out of the training set to build a development set. All items are re-sampled to 16\,kHz.

\subsection{Model configurations}
The models are constructed with 16 flow blocks and a subsampling factor $G=12$. The subnetwork has 8 layers of dilated convolutions implemented as depthwise separable convolutions \cite{depthwise}. In contrast to \cite{strauss2021}, the output channels of the dilated convolutions are set to 128 and the conditional input layer is replaced by a depthwise separable convolution. This configuration of the model using the time domain input has a total of 8.8\,M parameters, which is significantly lower than the one in \cite{strauss2021}. Using the double coupling scheme approximately doubles the amount of parameters, since two subnetworks have to be learned for each block. From each training audio file, 1\,s long chunks are randomly extracted and given as inputs to the network. In enhancement the entire signal is processed at once. The models are trained with a batch size of 4 and Adam optimizer (learning rate $=0.001$). The learning rate is decayed by a factor of $0.5$, if the validation loss did not decrease for 10 consecutive epochs. All models are trained for 200 epochs.
Similar to previous works \cite{Prenger2019, strauss2021}, using a lower standard deviation for the sampling distribution in inference showed a slightly better performance experimentally. Hence, the standard deviation was lowered from $\sigma=1.0$ in training to $\sigma=0.9$ in enhancement. 
For the Mel-spectrogram the FFT parameters are: 512 samples for the input and window size, Hann window, and 75\% overlap. The spectrogram includes 80 frequency bands.
The APG is implemented with a filter order of 4 and a lookahead factor for group delay compensation of 0.7. The minimum center frequency is set to 40\,Hz with a total of 80 frequency bands and the maximum center frequency just below Nyquist frequency.

%\subsection{Ablation study}
%We conducted a small ablation study to verify that the architectural extensions lead to an increased performance.

\section{Evaluation}
We compare the following flow-based systems. The original model with single coupling and time domain conditional input \cite{strauss2021} is denoted as \textbf{SE-Flow$_\text{sc}$}, while the proposed double coupling version is referred to with \textbf{SE-Flow}. The varied input conditions are characterized by the suffix \mbox{\textbf{-Mel}}, and \textbf{-APG}. Two state-of-the-art generative models are also considered, namely the MetricGAN+ \cite{fu21_interspeech}, and \mbox{CDiffuSE}~\cite{lu2022conditional}. The samples enhanced by MetricGAN+ are obtained from the model in the speechbrain project \cite{speechbrain}. For \mbox{CDiffuSE}, the outputs were kindly provided by the authors of the corresponding paper.

\subsection{Computational evaluation metrics}    
The methods are first evaluated with computational metrics.
PESQ \cite{pesq} (worst: -0.5; best: 4.5) and the mean opinion score estimating composite measures \cite{Hu2008} (worst: 1; best: 5) are commonly reported on this dataset. For further insights, STOI \cite{stoi} (worst: 0; best: 1) and the 2f-model score \cite{2f, Torcoli21-TASLP} (worst: 0; best: 100) are also reported.

\subsection{Listening test}
The considered methods are also compared via a listening test following the MUSHRA methodology~\cite{mushra2015} with a reference and a 3.5\,kHz low-pass anchor. The participants were instructed to rate the \textit{overall sound quality} of the presented items with regard to the reference.
The test items were selected from the BUS and CAFE noise settings and only the most difficult SNR conditions, i.e., 2.5 and 7.5\,dB SNR.  Per test speaker, one item of at least 3\,s was randomly selected for a total of 8 items. The computational evaluation was repeated on the test items and confirmed that this item selection was not biased towards a particular model. The samples used for this test along with the unprocessed input signals can be found online\footnote{\url{https://www.audiolabs-erlangen.de/resources/2022-SLT-improved_SE_Flow}}.

The raw outputs from the different systems differ greatly in overall energy. For one example item, output integrated loudness \cite{loudness_standart} can range from -17 to -24 loudness units full scale (LUFS), while both noisy input and clean speech are -22.8\,LUFS. Moreover, different levels of leaking noise are observed after processing. This can make it very difficult to assign an overall quality score to the compared systems, as noise suppression and speech quality are often inversely proportional.
In order to ensure that a fair comparison of the systems is possible, leaking noise level matching and loudness normalization are carried out, similarly to \cite{strauss21_interspeech}. First, \textit{background} components are obtained by subtracting the enhanced output or the clean reference from the input mixture. Then:
\begin{enumerate}[topsep=2pt, partopsep=1pt, itemsep=1pt,parsep=2pt]
	
	\item The reference condition is created by mixing the reference clean speech with the corresponding background component, with an attenuation factor of 30\,dB.
	\item Speech activity information is determined by thresholding the envelope of the clean reference.
	\item The integrated loudness of the non-speech parts of the reference condition are determined (gating deactivated).
	\item For each test condition, the noise attenuation level is obtained iteratively, until the same loudness of the non-speech parts is reached as in the reference condition. The same speech activity information gathered for the reference condition is used.
	\item Each condition is normalized to -23\,LUFS (integrated loudness, gating deactivated).
\end{enumerate}

The test was conducted online using webMUSHRA \cite{webmushra} with each participant using their own PC and headphones. The participants were 20 fellow colleagues with various level of experience in audio research. No results had to be removed in accordance to the MUSHRA post-screening procedure.
Note that since we want to make sure that the amount of leaking noise is comparable across systems, the test items for CDiffuSE are generated from the raw network output, while the numbers reported in the paper include a recombination with the original noisy signal. 

\section{Results and Discussion}
%\vspace{-0.2cm}
%\setlength{\tabcolsep}{8pt}\textbf{}
\begin{table}[htb]
	\setlength{\tabcolsep}{3.2pt}
	\centering
	\caption{Computational evaluation results for the test set (VoiceBank-DEMAND). 
		%SE-Flow without suffix refers to the time domain models. Suffix \textbf{-Mel} and \textbf{-APG} are the corresponding Mel-spectrogram and APG versions. The subscript \textbf{sc} indicates the original single coupling architecture. 
		Mean values. Best results in bold.}
	\resizebox{0.47\textwidth}{!}{%
		\begin{tabular}{lcccccc}\toprule
			Method     & \multicolumn{1}{r}{PESQ} & \multicolumn{1}{r}{CSIG} & \multicolumn{1}{r}{CBAK} & \multicolumn{1}{r}{COVL} & \multicolumn{1}{r}{STOI}  & \multicolumn{1}{r}{2f-model} \\\midrule % {[}\%{]}}
		Noisy      & 1.97                     & 3.35                     & 2.45                     & 2.63                     & 0.92                                & 31.70                         \\
		CDiffuSE \cite{lu2022conditional} 	&	2.52			& 3.72					& 2.91				& 3.10 			& 0.91		&  33.65 \\
		MetricGAN+ \cite{fu21_interspeech} & \textbf{3.13}                     & \textbf{4.08}                     & \textbf{3.16}                     & \textbf{3.60}                     & \textbf{0.93}                                & 34.36                         \\\midrule
		SE-Flow$_\text{sc} $   &  2.24   &   3.60  &  2.95    &  2.91   &      0.90        &   44.47    \\ 
		SE-Flow    &  2.41   &   3.79  &  3.11    &  3.09   &      \textbf{0.93}        &   \textbf{46.92}    \\ 
		SE-Flow-Mel    & 1.63     & 2.84     & 2.00    & 2.19     & 0.85              & 35.63        \\ 
		SE-Flow-APG    & 2.05    & 3.30     & 2.51     & 2.65    & 0.89              & 40.16       \\ \bottomrule
\end{tabular}}
%\vspace{-0.5cm}
\label{tab:results}
\end{table}

\subsection{Computational evaluation results}

Table~\ref{tab:results} shows the results of the computational evaluation on the test set. The computational metrics were also evaluated separately for the conditions 7.5 and 2.5\,dB input SNR, as well as considering exclusively the items selected for the listening test. While the metrics show significantly different absolute values, the main trends and the ranking of the methods remain the same as in Table~\ref{tab:results}.

The metrics show that SE-Flow outperforms the single coupling version, confirming the benefits of the proposed double coupling architecture at the cost of a more complex network. \mbox{SE-Flow} with time domain conditional input shows the best performance among the flow-based models. SE-Flow-Mel shows the lowest performance with some metrics even worse than the noisy baseline. %This suggests that the model introduces significant artifacts in the processed outcome. 
One possible explanation is that the Mel-representation of the noisy speech is sub-optimal for the application at hand, possibly because it does not provide phase information about the input. 
%One possible explanation is, that the transformation of a noisy speech utterance to a Mel-representation  does not deliver sufficiently enough information about the underlying clean signal. Another possibility is, that the algorithm has no information about the phase of the input signal in the enhancement step. 
SE-Flow-APG exhibits lower results in PESQ and in the composite measures than the other methods, but its 2f-model results only lay behind the time domain flow models. 
%Looking at the results of SE-Flow-APG it can be seen that, while the model has lower results in PESQ and the composite measures compared to other methods, the 2-f model results only lay behind the time domain flow model. 

MetricGAN+ shows the best results in PESQ and the composite measures. These results are somewhat to be expected since MetricGAN+ directly optimizes PESQ. 
In terms of STOI, MetricGAN+ shows the best performance together with \mbox{SE-Flow}. CDiffuSE outperforms all flow-based models in terms of PESQ, and composite measures, only staying behind MetricGAN+. This confirms the results reported in their publication with regard to other time-domain generative models.  With regard to the 2f-model, the time domain SE-Flow shows the best performance among all methods.

\subsection{Listening test results}

\begin{figure}[tb]
\resizebox{0.48\textwidth}{!}{%
	\centering
	\begin{minipage}[b]{\textwidth}% 0.88
		\includegraphics[width=1\textwidth]{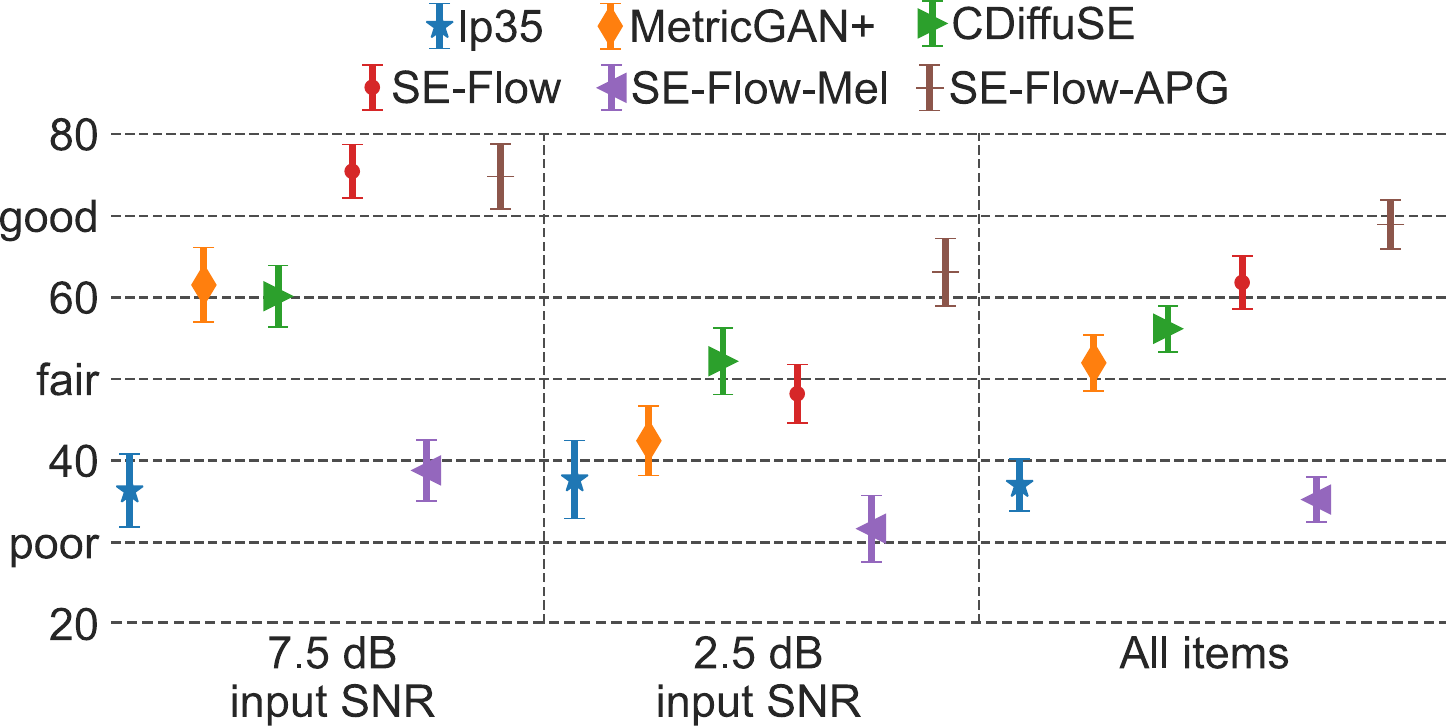}% mushra
\end{minipage}}
\caption{Listening test results (20 listeners). Mean values and 95\% confidence intervals (Student's t-distribution). The results are shown for the different input conditions (7.5\,dB and 2.5\,dB SNR) and over all items in the test. The hidden reference (not shown here) was always rated $>=$ 95.
	%Next to the tested systems \textbf{hiddenRef} described the reference condition and \textbf{lp35} the low anchor condition obtained by a 3.5\,kHz low-pass filter.
	(Best viewed in colors)}
% \vspace{-0.82cm}
\label{fig:mushra}
\end{figure}

The results of the listening test are depicted in Figure~\ref{fig:mushra}. The average results over all items show that SE-Flow-APG performs best among all methods, being rated as having \textit{good} quality on average, and with the confidence intervals not overlapping with the other methods. SE-Flow follows. Despite the high values in the computational evaluation metrics, MetricGAN+ performs worse than SE-Flow, SE-Flow-APG and CDiffuSE. Inspecting some of the enhanced samples reveal artifacts in the voice timbre, which could explain the low results in part.  CDiffuSE takes the third place behind both suggested flow-based approaches. Examining the respective samples reveals a low-pass-filter-like characteristic of the output samples, which could explain the results to some extend. As indicated also by the computational evaluation, \mbox{SE-Flow-Mel} exhibits the worst performance, with scores similar to the low-pass anchor. \\
\indent Considering the results grouped by input SNR condition, SE-Flow-APG performs best at the lowest SNR condition, while being on par with SE-Flow at 7.5\,dB SNR.	In fact, the performance of SE-Flow drops dramatically going from 7.5 to 2.5\,dB SNR, where the superiority of SE-Flow-APG is evident.  Also, MetricGAN+ is close to the \textit{good} quality range for the higher SNR condition, but it drops 15 MUSHRA points when tested at 2.5\,dB SNR. CDiffuSE shows more robustness across SNRs, but overall lower quality than SE-Flow-APG.\\
\indent It is worth highlighting that the results from the listening test are in partial disagreement with the results from the computational metrics. In fact, even if computational metrics can be extremely useful for their convenience and reproducibility, their correlation with perceived audio quality is often low~\cite{Torcoli21-TASLP}. For this reason, conclusions drawn exclusively from computational metrics should be taken with care, as they can be partially misleading in terms of perceived quality. 

\section{Conclusion}
In this paper, several improvements to a flow-based SE model are introduced. With the presented double coupling scheme, the model  processes the entire input signal in each coupling layer, leading to higher capacity and performance. Additional experiments consider different representations for the conditional input.  Despite being a common choice in related fields, Mel-spectrograms proved not to be a suitable choice for flow-based SE. As an alternative, a proposed Bark spaced All-Pole Gammatone filterbank-based pre-processing with increased time resolution overcomes the Mel induced problems. While the results of popular computational metrics are behind state-of-the-art generative models, the outcome of a listening test indicates, that flow-based SE using a time-domain or gammatone-filtered conditional signal has favourable perceptual performance. Hereby, it was shown that the proposed method not only outperforms the compared generative models, but the performance also remains strong throughout different SNR conditions. 
 
\section{ACKNOWLEDGMENTS}
\label{sec:ack}

The authors would like to thank Yen-Ju Lu and Yu Tsao for
sharing the test samples of their diffusion model for comparison.

\bibliographystyle{IEEEbib}
\bibliography{strings,refs}

\end{document}